\documentstyle[aps,preprint]{revtex}
\begin{document}
\tighten
\title{\bf Non-perturbative evolution equations for 
the tricritical theory}

\author{Flavio S. Nogueira\footnote[1]{e-mail: 
nogueira@orphee.polytechnique.fr}}

\address{Centre de Physique Th\'eorique, Ecole Polytechnique F91128 
Palaiseau Cedex, FRANCE \\
and \\
Instituto de Matem\'atica e Estat\'{\i}stica - 
Universidade do Estado do Rio de Janeiro, Rua S. Francisco 
Xavier 524, Bl. B, Rio de Janeiro, RJ 20559-900, BRAZIL}

\date{Received \today}

\maketitle

\begin{abstract}
The $N$ component scalar tricritical theory is considered in a 
non-perturbative setting. We derive non-perturbative beta 
functions for the relevant couplings in $d\leq 3$. The beta 
functions are obtained through the use of an exact evolution 
equation for the so called effective average action. In 
$d=3$ it is 
established the existence of an ultraviolet stable fixed point 
for $N>4$. This confirms earlier results obtained using the  
$1/N$ expansion where such a fixed point is believed to exist 
at least for $N\gtrsim 1000$.
\end{abstract}
\pacs{  }

The theory of tricritical phenomena is very important for  
the understanding of the statistical mechanics of binary mixtures 
such as $He_{3}-He_{4}$ \cite{Griffiths}. The tricritical point in the 
phase diagram corresponds to a regime where three phases coexist. 
The statistical mechanics of tricritical phenomena can be investigated 
through an appropriate Ginzburg-Landau action functional. The tricritical 
exponents are then determined by computing the fluctuations around the 
corresponding mean field behavior \cite{Lawrie}. This can be done by 
standard $\epsilon$-expansion techniques \cite{WK}. However, this procedure  imply the existence of a non-trivial infrared stable fixed 
point and a trivial ultraviolet fixed point for 
$\epsilon\neq 0$. For $d=3$ there is only a 
Gaussian fixed point. Thus, in the framework of traditional 
$\epsilon$-expansion, the tricritical theory behaves exactly in 
the same way as a critical theory \cite{ItDr,Aragao}. 
Nevertheless, critical theories remain to exhibit this same behavior 
for arbitrary number of field components, even in the context of 
a $1/N$ expansion 
\cite{Zinn-Justin}. On the other hand, this is not true for tricritical 
theories. Indeed, large $N$ results points for the existence of a
non-trivial  
ultraviolet stable fixed point in 
$d=3$ \cite{Pisarski}. This result is believed 
to be true at least for $N\gtrsim 1000$.

In this paper we study the tricritical theory  non-perturbatively for 
arbitrary values of $N$. Our starting point is an exact evolution 
equation for the quantum effective action at a $k$ scale, the 
average action $\Gamma_{k}$ \cite{Wetterich1}. The exact evolution 
equation for the average action is an exact renormalization group 
equation where the fast modes are effectively integrated out. It is 
closely related to other versions of exact renormalization group 
equations \cite{WK,Wegner}. The average action interpolates between 
the classical action and the usual quantum effective action. 
Specifically, we have $\Gamma_{k}\rightarrow S$ as $k\rightarrow\Lambda$, 
$S$ being the classical action and $\Lambda$ an ultraviolet cutoff, and 
$\Gamma_{k}\rightarrow\Gamma$ as $k\rightarrow 0$, where $\Gamma$ is the 
usual quantum effective action. The exact renormalization group equation 
is given by \cite{Wetterich1,Grater}

\begin{equation}
\label{exac}
k\frac{\partial\Gamma_{k}}{\partial k}=\frac{1}{2}Tr\left[
(\Gamma^{(2)}_{k}+R_{k})^{-1}k\frac{\partial R_{k}}{\partial k}\right].
\end{equation}
In above, $R_{k}$ is a suitable smooth infrared cutoff function whose 
definition is not unique. $\Gamma^{(2)}_{k}$ is the second functional 
derivative of $\Gamma_{k}$ with respect to the fields. The trace is 
meant $Tr=\Omega\sum_{a}\int d^{d}q/(2\pi)^{d}$, where $\Omega$ is 
the volume of the system and the sum is over the color indices. A 
choice for $R_{k}$ frequently used in the literature is \cite{Wetterich2}

\begin{equation}
R_{k}(q)=Z_{k}q^{2}f_{k}(q),
\end{equation}
where

\begin{equation}
f_{k}(q)=\frac{1}{exp(q^{2}/k^{2})-1}.
\end{equation}
$Z_{k}$ is a suitable wave function renormalization which generally is a 
function of the momentum $q$ and $\rho=\phi_{a}\phi_{a}/2$ for theories 
with a $O(N)$ symmetry. However, we will use in this paper an 
approximation for which $Z_{k}$ is uniform. From $Z_{k}$ we define 
the anomalous dimension by 

\begin{equation}
\eta=-k\frac{\partial\ln Z_{k}}{\partial k}.
\end{equation}

A thorough study of critical theories using Eq.(\ref{exac}) was made 
by Tetradis and Wetterich \cite{Wetterich2}. The Kosterlitz-Thouless 
phase transition was studied through the same formalism by 
Gr\"ater and Wetterich \cite{Grater}. The tricritical theory can be 
studied with the same formalism. The following truncation will be 
used for the average action:

\begin{equation}
\Gamma_{k}=\int d^{d}x\left[\frac{1}{2}Z_{k}(\rho,-\partial_{\mu}
\partial_{\mu})\partial_{\mu}\phi_{a}\partial_{\mu}\phi_{a}+
U_{k}(\rho)\right].
\end{equation}
Let us consider a 
uniform background field configuration of the form $\phi_{a}=\delta_{a1}\phi$ 
and assume that the effective potential $U_{k}(\rho)$ can be expanded 
around a broken symmetry 
minimum $\rho_{0}(k)=\omega_{k}$. The critical theory is defined 
through the condition $U'_{k}(\omega_{k})=0$, where the prime means a 
derivative with respect to $\rho$. In this situation and trucating the 
potential up to second order in $\rho$, we get the non-perturbative 
evolution equations obtained by Gr\"ater and Wetterich \cite{Grater} 
(See also ref. \cite{Wetterich2}). The tricritical theory is defined 
through the condition $U'_{k}(\omega_{k})=U''_{k}(\omega_{k})=0$. Let 
us assume the following form for the effective potential:

\begin{equation}
U_{k}(\rho)=U_{k}(\omega_{k})+\frac{U'''_{k}(\omega_{k})}{3!}
(\rho-\omega_{k})^{3}.
\end{equation}
Let us define the dimensionless couplings 
$\tilde{u}_{k}$, $\tilde{\omega}_{k}$ and $\tilde{g}_{k}$ by 
$U_{k}(\omega_{k})=k^{d}\tilde{u}_{k}$, 
$\omega_{k}=Z_{k}^{-1}k^{d-2}\tilde{\omega}_{k}$ and 
$U'''_{k}(\omega_{k})=Z_{k}^{3}k^{6-2d}\tilde{g}_{k}$. The evolution 
equations for the dimensionless couplings are given by

\begin{eqnarray}
\label{b1}
k\frac{\partial\tilde{u}_{k}}{\partial k}&=&-d\tilde{u}_{k}+\frac{S_{d}N}{4}[I_{1}^{d}-I_{2}^{d}-\eta(
J_{1}^{d}-J_{2}^{d})]+\frac{S_{d}N}{4}\tilde{g}_{k}\tilde{\omega}_{k}^{2}
(I_{2}^{d}-\eta J_{2}^{d})\nonumber \\
&+&S_{d}\tilde{\omega}_{k}^{4}\tilde{g}_{k}^{2}[I_{2}^{d}-I_{3}^{d}-\eta(
J_{2}^{d}-J_{3}^{d})],
\end{eqnarray}

\begin{eqnarray}
\label{b2}
k\frac{\partial\tilde\omega_{k}}{\partial k}&=&(2-d-\eta)\tilde{\omega}_{k}+
\frac{S_{d}N}{2}(I_{2}^{d}-\eta J_{2}^{d})\nonumber \\
&-&2S_{d}\tilde{\omega}_{k}^{2}\tilde{g}_{k}(I_{3}^{d}-\eta J_{3}^{d}),
\end{eqnarray}

\begin{eqnarray}
\label{b3}
k\frac{\partial\tilde{g}_{k}}{\partial k}&=&(2d-6+3\eta)\tilde{g}_{k}
+15S_{d}\tilde{\omega}_{k}\tilde{g}_{k}^{2}(I_{3}^{d}-\eta J_{3}^{d})
\nonumber \\
&-&12S_{d}\tilde{\omega}_{k}^{3}\tilde{g}_{k}^{3}(I_{4}^{d}-\eta J_{4}^{d}),
\end{eqnarray}
where 

\begin{equation}
I_{n}^{d}=
\int_{0}^{\infty}dx\frac{x^{\frac{d}{2}-n}H(x)}{[1+F(x)]^{n}}, 
\end{equation}
and
\begin{equation}
J_{n}^{d}=\int_{0}^{\infty}dx\frac{x^{\frac{d}{2}-n}F(x)}{[1+F(x)]^{n}},
\end{equation} 
with $F(x)=1/(e^{x}-1)$, $H(x)=e^{x}[F(x)]^{2}$ and 
$S_{d}=2^{1-d}\pi^{-d/2}/\Gamma(d/2)$. Note that the above result differs 
considerably from that one obtained by standard perturbative methods. 

It should be observed that the anomalous dimension $\eta$ is not known 
{\it a priori}. However, it can be calculated from the evolution 
equation for $\Gamma^{2}_{k}$, which is readily obtained from Eq.(\ref{exac}). 
The lenghty computation of $\eta$ for a critical theory is described in 
detail in reference \cite{Wetterich2}. Here we have a simplification due to 
the fact that we are assuming $U''_{k}(\omega_{k})=0$. Moreover, if we assume 
further a truncation such that $Z'_{k}(\omega_{k})=Z''_{k}(\omega_{k})=0$, we 
get the simple result $\eta=0$ for the anomalous dimension. Again, this 
result is different from the perturbative one, where it is found that 
$\eta\sim\tilde{g}^{2}_{k}$. The result $\eta=0$ is consistent with the 
truncation and the hypothesis of a $\rho$-independent wave function 
renormalization.

Given that $\eta=0$, we have for any dimension $d>2$ the following fixed point:

\begin{equation}
\label{inf1}
\tilde{u}_{k}^{*}=\frac{S_{d}N}{4d}(I_{1}^{d}-I_{2}^{d}),
\end{equation}

\begin{equation}
\label{inf2}
\tilde{\omega}^{*}_{k}=\frac{S_{d}NI_{2}^{d}}{2(d-2)},
\end{equation}

\begin{equation}
\label{inf3}
\tilde{g}^{*}_{k}=0.
\end{equation}
This fixed point is infrared stable in the direction along the 
$\tilde{g}_{k}$-axis. Along the $\tilde{\omega}_{k}$-axis it is infrared 
repulsive (ultraviolet attractive).
For $d=3$ we have also a non-trivial ultraviolet fixed point, provided 
$N>4$:

\begin{equation}
\label{UV2}
\tilde{u}_{k}^{*}=\frac{1}{2d\pi^{2}}\left\{\frac{N}{4}[I_{1}^{3}-I_{2}^{3}]+
\frac{N}{4}I_{2}^{3}\tilde{g}_{k}^{*}\tilde{\omega}_{k}^{*2}+
\tilde{\omega}_{k}^{*4}\tilde{g}_{k}^{*2}[I_{2}^{3}-I_{3}^{3}]
\right\},
\end{equation}

\begin{equation}
\label{UV3}
\tilde{\omega}_{k}^{*}=\frac{1}{4\pi^{2}}\left[NI_{2}^{3}-
\frac{5(I_{3}^{3})^{2}}{I_{4}^{3}}\right],
\end{equation}

\begin{equation}
\label{UV4}
\tilde{g}_{k}^{*}=\frac{20\pi^{4}I_{3}^{3}}{I_{4}^{3}\left[NI_{2}^{3}-
\frac{5(I_{3}^{3})^{2}}{I_{4}^{3}}\right]^{2}}.
\end{equation}
The existence of the above fixed point is a remarkable result. This result 
was known before only for sufficiently large values of $N$ \cite{Pisarski}. 
For instance, the large $N$ result is assumed to be consistent for 
$N\gtrsim 1000$. 
We have established the same result for a much smaller value of 
$N$. For $N\leq 4$ we have that the ultraviolet fixed point becomes 
physically inaccessible due to the fact that $\tilde{\omega}_{k}^{*}<0$ 
in this case.

Let us consider what happens if we suppose $N$ large in the present 
non-perturbative setting. For $N\rightarrow\infty$ and 
$d=3$ we have that the 
ultraviolet fixed point tends to the fixed point given by 
Eqs.(\ref{inf1},\ref{inf2},\ref{inf3}). In this limit 
$\omega_{k}\sim O(N)$ and it is convenient to 
rescale $g_{k}\rightarrow g_{k}/N^{2}$. The flow equations are solved exactly 
in this limit and we have

\begin{equation}
\tilde{\omega}_{k}=\frac{\Lambda}{k}\tilde{\omega}_{\Lambda}+
\frac{I_{2}^{3}}{4\pi^{2}}\ln(\frac{k}{\Lambda}),
\end{equation}

\begin{equation}
\tilde{g}_{k}=\tilde{g}_{\Lambda}.
\end{equation}
One obtains the familiar result that $\tilde{g}$ gets unrenormalized when 
$N\rightarrow\infty$ \cite{Moshe}.

The $d=2$ case is not accessible directly because $d=2$ is a pole for 
Eq.(\ref{inf2}). Assume, however, that it is plausible to write 
$d=3-\epsilon$. It is not necessary to manipulate $I_{2}^{d}$ 
with dimensional regularization because $I_{2}^{2}=1$ exactly. Then, the 
fixed point (\ref{inf2}) can be written up to any order in $\epsilon$. The 
problem is that $I_{1}^{d}$ is infrared divergent for $d=2$ and Eq.(\ref{b1}) 
makes no sense. The trouble here is due to the fact that the truncation which 
leads to $\eta=0$ is not reliable for the $d=2$ situation where the 
infrared divergences are very severe. The calculation of $\eta$ must be 
considerably improved in order to get consistent results. It is worth to mention that, under the same hypothesis for the wave function renormalization 
$Z_{k}$, the critical theory has accessible fixed points in $d=2$ 
\cite{Grater}. This is due to the fact that $U'_{k}(\omega_{k})\neq 0$ in this 
case. This leads to a nonzero value of $\eta$ implying in turn a better 
control of the infrared behavior. The authors of reference \cite{Grater} 
were successful in describing the $d=2$ physics including the 
Kosterlitz-Thouless phase transition. They study the critical behavior 
in a scenario consistent with the Mermim-Wagner theorem. 

In summary, we obtained the non-perturbative flow equations for the 
tricritical field theory. We established the existence of a non-trivial 
ultraviolet fixed point for $N>4$ and $d=3$. This confirms early results 
obtained through the $1/N$ expansion formalism. The main difference is 
that we do not need a very large $N$ in the present situation. An 
interesting question concerns the improvement of the computation of 
the anomalous dimension.

\section*{Acknowledments}

The author would like to thank Prof. N. F. Svaiter for discussions and 
advice and Prof. A. P. C. Malbouisson for his encouragement. The author would 
like to thank 
also Prof. V. Rivasseau for the hospitality in the Centre de 
Physique Th\'eorique, Ecole Polytechnique. This 
work was supported in part by the brazilian agency CNPq.

\end{document}